\documentclass[runningheads]{llncs}
\usepackage{graphicx}
\usepackage{xcolor}
\usepackage{array}
\usepackage{booktabs, multirow} % for borders and merged ranges
\usepackage{soul}% for underlines
\usepackage{changepage}
\usepackage{threeparttable}% for wide tables
\usepackage{comment}
\usepackage[hidelinks]{hyperref}

\newcommand{\DOI}[1]{\textbf{DOI}: #1}

\begin{document}
\title{ChatGPT as a mapping assistant: A novel method to enrich maps with generative AI and content derived from street-level photographs}
\titlerunning{SDSS 2023}

% IT's just the order I reached out to you, we can change it
\author{Levente Juhász\inst{1}\orcidID{0000-0003-3393-4021} \and
Peter Mooney\inst{2}\orcidID{0000-0002-2389-3783} \and
Hartwig H. Hochmair\inst{3}\orcidID{0000-0002-7064-8238} \and
Boyuan Guan\inst{1}}
\authorrunning{L. Juhász et al.}
\institute{GIS Center, Florida International University, Miami, FL 33199, USA \email{\{ljuhasz,bguan\}@fiu.edu} \and
Department of Computer Science, Maynooth University, Co. Kildare, Ireland\\
\email{peter.mooney@mu.ie} \and
Geomatics Sciences, University of Florida, Ft. Lauderdale, FL 33144, USA\\
\email{hhhochmair@ufl.edu}}
\maketitle              % typeset the header of the contribution
\begin{abstract}
 This paper explores the concept of leveraging generative AI as a mapping assistant for enhancing the efficiency of collaborative mapping. We present the results of an experiment that combines multiple sources of volunteered geographic information (VGI) and large language models (LLMs). Three analysts described the content of crowdsourced Mapillary street-level photographs taken along roads in a small test area in Miami, Florida. \texttt{GPT-3.5-turbo} was instructed to suggest the most appropriate tagging for each road in OpenStreetMap (OSM). The study also explores the utilization of BLIP-2, a state-of-the-art multimodal pre-training method as an artificial analyst of street-level photographs in addition to human analysts. Results demonstrate two ways to effectively increase the accuracy of mapping suggestions without modifying the underlying AI models: by (1) providing a more detailed description of source photographs, and (2) combining prompt engineering with additional context (e.g. location and objects detected along a road). The first approach increases the accuracy of the suggestion by up to 29\%, and the second one by up to 20\%.

\end{abstract}
\DOI 10.25436/E2ZW27
\let\thefootnote\relax\footnotetext{\textbf{Cite as:} Juhász, L., Mooney, P., Hochmair, H.H., Guan, B. (2023). ChatGPT as a mapping assistant: A novel method to enrich maps with generative AI and content derived from street-level photographs. \textit{Spatial Data Science Symposium 2023 Paper Proceedings}. UC Santa Barbara Center for Spatial Studies.  \url{https://doi.org/10.25436/E2ZW27}}

\section{Introduction}\label{intro}

\begin{figure}
\includegraphics[width=\textwidth]{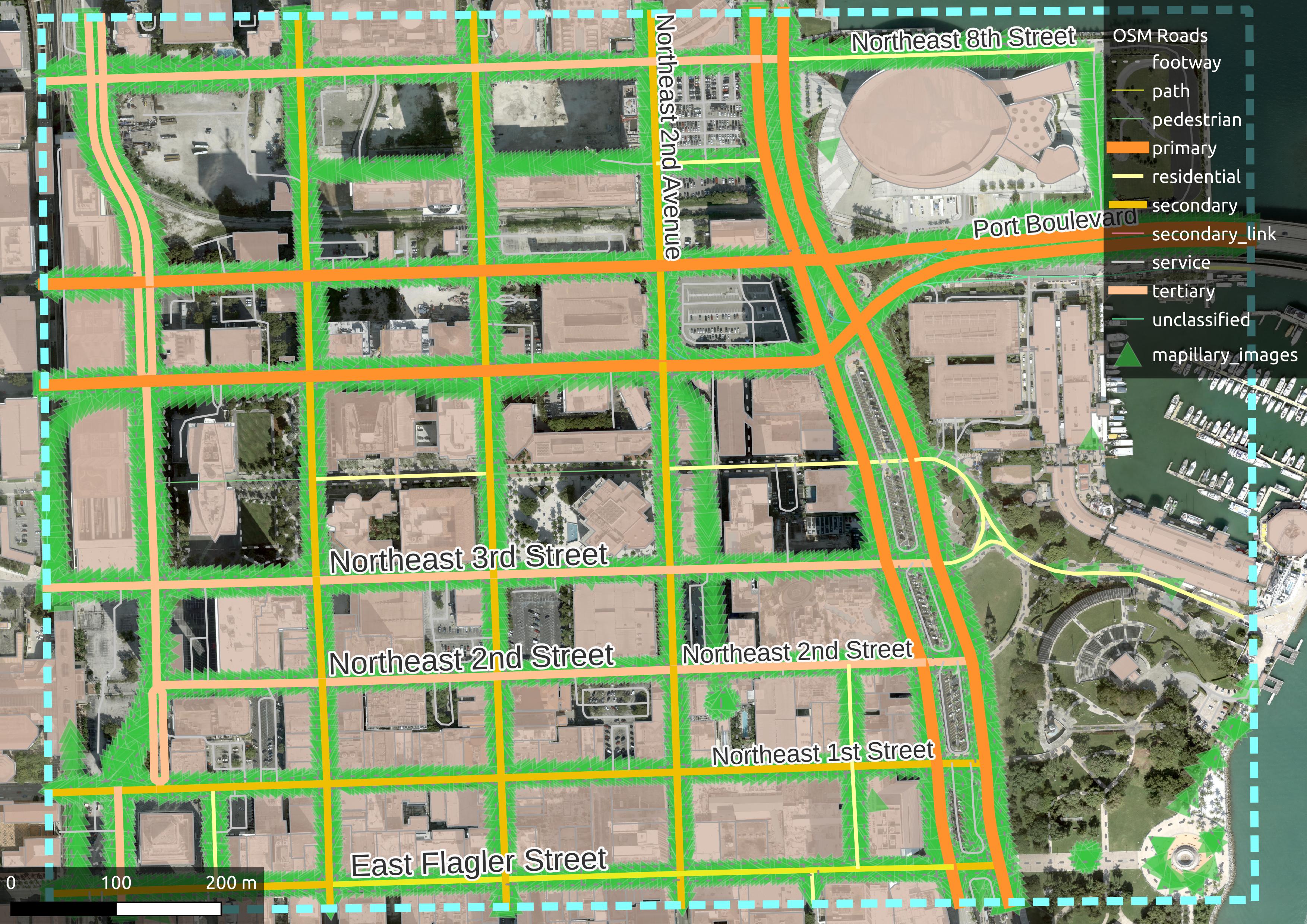}
\caption{OpenStreetMap roads and Mapillary images in the study area near Downtown Miami} \label{fig1}
\end{figure}

Generative artificial intelligence (AI) is a type of AI that can produce various types of content, including text, images, audio, code, and simulations. It has gained enormous attention since the public release of ChatGPT in late 2022. ChatGPT is an example of a Large Language Model (LLM), which is a form of generative AI that produces human-like language. Since the launch of ChatGPT, researchers, including the geographic information science (GIScience) community, have been trying to understand the potential role of AI for research, teaching, and applications. ChatGPT can be used extensively for Natural Language Processing (NLP) tasks such as text generation, language translation, writing software code, and generating answers to a plethora of questions, engendering both positive and adverse impacts~\cite{DWIVEDI2023102642}. The emergence of generative AI has introduced transformative opportunities for spatial data science. In this paper, we explore the potential of generative AI to assist human cartographers and GIS professionals in increasing the quality of maps, using OSM as a test case (Figure \ref{fig1}). 

GeoAI has been part of the GIScience discourse in recent years. For example, Janowicz et al. \cite{janowiczetaleditorial} elaborated on whether it was possible to develop an artificial GIS analyst that passes a domain specific Turing test. Although these questions are still largely unanswered, utilizing LLMs and foundational models in geospatial contexts contributes to this direction. Despite the challenges due to the different nature of LLM training methodologies and human learning of spatial concepts \cite{mooney_towards_2023}, these tools and methods are being explored, for example to generate maps \cite{ijgi12070284}. Our study fits into this direction by using an LLM (ChatGPT) and a multimodal pre-training method (BLIP-2) to connect visual and language information in the context of mapping. We explore the larger question of whether \textit{generative AI is a useful tool in the context of creating and enriching map databases} and more specifically investigate the following research questions:

\begin{enumerate}
    \item Is generative AI capable of turning natural language text descriptions into the correct attribute tagging of road features in digital maps?
    \item For this problem, can the accuracy of suggestions be improved through prompt engineering \cite{reynolds_prompt_2021}?
    \item To what extent can the work of human analysts be substituted with generative AI approaches within these types of mapping processes?
\end{enumerate}

Furthermore, our approach focuses on the fusion of freely available volunteered geographic information (VGI) \cite{goodchild_citizens_2007} data sources (OSM, Mapillary) and off-the-shelf AI tools to present a potentially low-cost and uniformly available solution. OSM is a collaborative project that aims to create a freely accessible worldwide map database (\url{https://openstreetmap.org}), while Mapillary crowdsources street-level photographs (\url{https://mapillary.com}) that power mapping and other applications, such as object detection, semantic segmentation and other computer vision algorithms to extract semantic information from imagery \cite{ertler_mapillary_2020}. While the use of VGI has not yet been explored in the context of generative AI, previous studies demonstrated the practicability of combining multiple sources of VGI to improve the mapping process~\cite{liu2021data}. More specifically, Mapillary street-level images are routinely used to enhance OSM \cite{juhasz_cross-linkage_2016,levente_juhasz_how_2017}.

\section{Study setup}\label{studysetup}

\subsection{Data sources and preparation}

In OSM, geographic features are annotated with key-value pairs to assign the correct feature category to them, a process called tagging \cite{mooney_annotation_2012}. For example, roads are assigned a \texttt{"highway"=<value>} tag where \texttt{<value>} indicates a specific road category, such as \texttt{"residential”} for a residential street.
 
OSM data is not homogeneous, and individual users may perceive roads differently and therefore assign different \texttt{"highway"} values to the same type of road. A list of \texttt{"highway"} tag values was established to better describe the meaning of each road category in OSM. Furthermore, the difference between some road categories, for example, primary and 'secondary, is more of an administrative nature rather than visual appearance. For example, a 2-lane road in rural areas could be considered primary, whereas a more heavily trafficked road in an urban environment might be categorized secondary. To consider semantic road categories rather than individual \texttt{"highway"} values as one of the evaluation methods, \texttt{"highway"} tag values representing similar roads in our dataset were grouped into four categories (Table \ref{tab1}).

\begin{table}
\caption{Grouping distinct \texttt{"highway"} tag values into semantically similar categories.}\label{tab1}
\begin{tabular}{|l|c|r|}
\hline
Category name &  OSM \texttt{"highway"} & \# of roads\\
\hline
Major, access controlled road &  \texttt{motorway|trunk} & 0\\
Main road &  \texttt{primary|secondary|tertiary} & 81\\
Regular road & \texttt{residential|unclassified|service} & 4\\
Not for motorized  traffic & \texttt{pedestrian|footway|cycleway} & 9\\
\hline
\end{tabular}
\end{table}

\begin{figure}
\includegraphics[width=\textwidth]{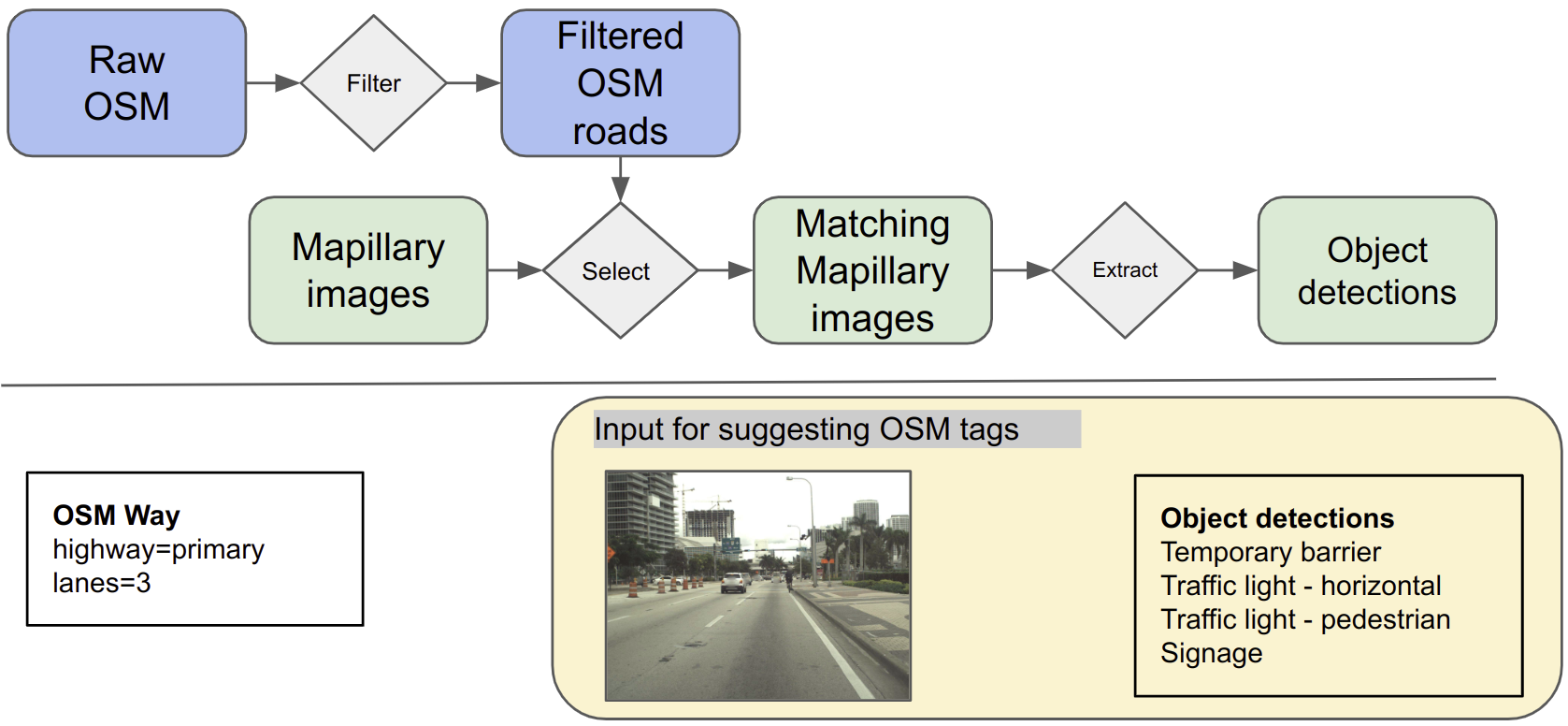}
\caption{Workflow for preparing input from Mapillary} \label{flowchart-mapillary}
\end{figure}

Figure \ref{flowchart-mapillary} shows the methodology to obtain OSM roads of interest with the corresponding Mapillary street-level images. First, all OSM roads with a \texttt{"highway"=*} tags were extracted within the study area. Then, short sections (\textless 50m), inaccessible roads, sidewalks along roadways and roads without street-level photo coverage were excluded. Retained OSM roads were matched with corresponding Mapillary photographs, so that each road segment would have at least one representative Mapillary image. Lastly, a list of objects detected in the corresponding image was also extracted from the Mapillary API. These inputs were further used as described in Section \ref{methodology}.

\subsection{Resources}\label{resources}

\subsubsection{AI tools and models}\label{aitools}

We utilize \texttt{GPT-3.5-turbo}, which is an advanced language model developed by OpenAI. It is an upgraded version of GPT-3, designed to offer improved performance and capabilities, and retains the large-scale architecture of its predecessor, enabling it to generate coherent and contextually relevant text \cite{ouyang_training_2022}. \texttt{GPT-3.5-turbo} serves as a powerful tool for natural language processing, content generation, and other language-related applications. In our study it is used to suggest OSM tagging based on pre-constructed prompts using the content of street-level images. The model was accessed through the OpenAI API.

BLIP-2 \cite{li_blip-2_2023} is a state-of-the-art, scalable multimodal pre-training method, designed to equip LLMs with the capability to understand images while keeping their parameters entirely frozen. This approach is built upon leveraging frozen pre-trained unimodal models and a proposed Querying Transformer (Q-Former), sequentially pre-trained for vision-language representation learning and vision-to-language generative learning. Despite operating with fewer trainable parameters, BLIP-2 has achieved exceptional performance in a variety of vision-language tasks and has shown potential for zero-shot image-to-text generation \cite{li_blip-2_2023}. The methodology proposed in BLIP-2 contributes towards the development of an advanced multimodal conversational AI agent. We leverage BLIP-2’s capability to generate image captions as well as to perform visual question-answering (Q\&A). A freely available sample implementation of BLIP-2 was used to conduct this experiment (\texttt{\url{https://replicate.com/andreasjansson/blip-2}}).

\subsubsection{Analysts}\label{analysts}

Three analysts were tasked to describe the visual content of street-level images (captioning), and to answer a few questions regarding the image content (visual Q\&A). Two (human) analysts were undergraduate students at Florida International University with previous GIS coursework. BLIP-2 was used to perform the same task as human analysts, and its responses were recorded as the third (artificial) analyst. Analysts were deliberately not given any guidelines as to how to describe images so that their answers would not be biased by prior knowledge about OSM and mapping. Table \ref{tab2} lists questions and tasks performed by analysts. 

\begin{table}
\caption{Questions and tasks performed by analysts.}\label{tab2}
\begin{tabular}{ | p{2cm} | p{6cm} | p{4cm} | }
\hline
\textbf{Variable} &  \textbf{Question/task} & \textbf{Example response}\\
\hline
\textit{\texttt{"caption"}} & Describe what you see in the photo in your own words. & \textit{A city road in an urban area along an elevated railway. There is a wide sidewalk on both sides and trees on the left.}\\
\hline
\textit{\texttt{"users"}} &  Who are the primary users of the road that is located in the middle of the photograph? Cars, pedestrians or bicyclists? & \textit{Cars}\\
\hline
\textit{\texttt{"lanes"}}& How many traffic lanes are there on the road that is in the middle of the photograph? & \textit{3}\\
\hline
\textit{\texttt{"surface"}} & What is the material of the surface of the road that is in the center of the photograph & \textit{Asphalt}\\
\hline
\textit{\texttt{"oneway"}} &Is the road that is in the center of the photograph one-way? & \textit{No}\\
\hline
\textit{\texttt{"lit"}} & Are there any street lights in the photograph? & \textit{Yes}\\
\hline
\end{tabular}
\end{table}

The answers of analysts differ in level of detail. For example, BLIP-2’s and Analyst \#2’s captions were significantly shorter on average (9 and 11 words, respectively) than Analyst \#1’s (37 words). BLIP-2’s responses were also found to be more generic (e.g. \texttt{"a city street with tall buildings in the background"}) than human analysts’. This allows us to explore the effect of providing increasing detail on tag suggestion accuracy.

\subsection{Methodology for suggesting OSM tags}\label{methodology}

\begin{figure}
\includegraphics[width=\textwidth]{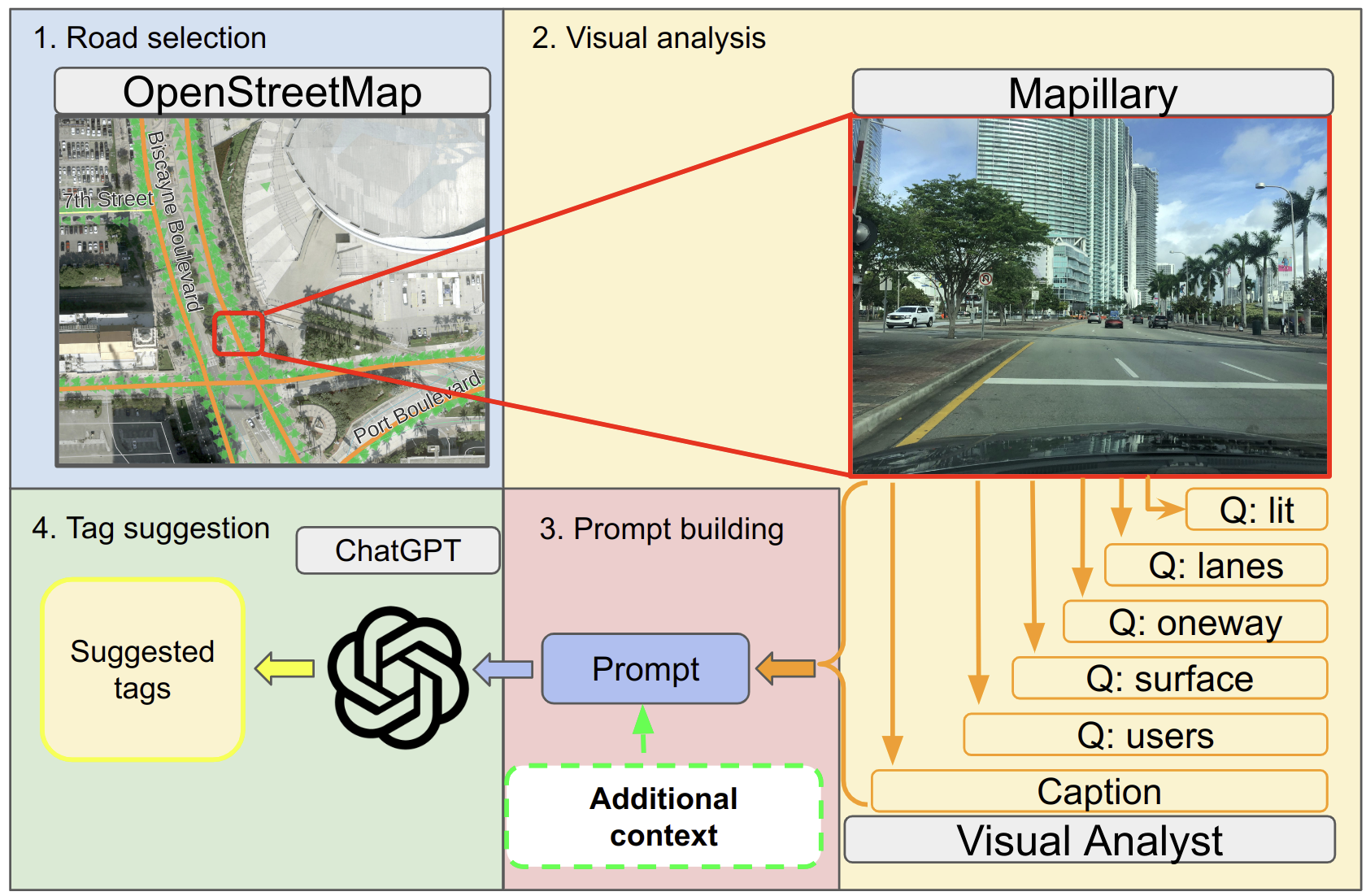}
\caption{Workflow of using ChatGPT to suggest OSM \texttt{"highway"} tags} \label{fig-method}
\end{figure}

Figure \ref{fig-method} shows the methodology for suggesting tags for an OSM road. For each retained road in the area, the corresponding Mapillary images were shown to analysts described in Section \ref{analysts}. Analysts created an image caption and answered simple questions as described in Table \ref{tab2}. These responses in combination with additional context were used to build prompts for an LLM to suggest OSM tags. To explore what influences the accuracy of suggested tags, a series of prompts were developed that differ in the level of detail that is presented to the LLM.

All prompts start with the following message that provides context and instructs the model about the expected output format.

\begin{quote}
    \textit{Based on the following context that was derived from a street-level photograph showing the street, recommend the most suitable tagging for an OpenStreetMap highway feature. Omit the ’oneway’ and ’lit’ tags if the answer to the corresponding questions is no or N/A. Format your suggested key-value pairs as a JSON. Your response should only contain this JSON.}
\end{quote}

The remainder of individual prompts is organized into four scenarios constructed from the responses of analysts and additional context. Example responses from analysts and additional context are highlighted in bold.

The \texttt{Baseline scenario} uses only responses from analysts, and contains the following text in addition to the common message above:

\begin{quote}
   \textit{The content of the photograph was described as follows: \textbf{A city road in an urban area along an elevated railway. There is a wide sidewalk on both sides and trees on the left.} The road is mainly used by: \textbf{cars.} The surface of the road is: \textbf{asphalt.}}
    
    \textit{When asked how many traffic lanes there are on the road, one would answer: \textbf{3.}}
    
    \textit{When asked if this street is a one-way road, one would answer: \textbf{No.}}
    
    \textit{When asked if there are any street lights in the photograph, one would answer: \textbf{Yes.}}
    \end{quote}

The \texttt{Locational context (LC) enhanced scenario} provides ChatGPT with additional locational context that describes where the roads in questions are located. In addition to the baseline message, it contains the following:

\begin{quote}
    \textit{\textbf{The photograph was taken near Downtown Miami, Florida.}}
\end{quote}

The \texttt{Object detection (OD) enhanced scenario} uses a list of detected objects in addition to the baseline:

\begin{quote}
    \textit{When guessing the correct category, consider that the following list of objects (separated by semicolon) are present in the photograph: \textbf{Temporary barrier; Traffic light - horizontal; Traffic light - pedestrian; Signage}}
\end{quote}

Finally, \texttt{Object detection and locational context (OD + LC)} are combined into a new scenario that supplies both additional contexts for the language model. 

The last step in the process is to supply the prompts described above to \texttt{GPT-3.5-turbo} (ChatGPT for simplicity). The model responds with a JSON document containing the suggested OSM tagging for the roadway, e.g. (\texttt{"highway"} \texttt{="primary", "lanes"= 3}), which can be compared to the original OSM tags of the same roadway.

The final dataset contains 94 OSM highway features and their original tags. For four scenarios and three analysts described above, ChatGPT recommendations based on the corresponding prompts were also recorded for the same roadway, resulting in a total of 12 tagging suggestions. These suggestions are then compared to the original OSM tags to assess the accuracy of a particular scenario and analyst.

\section{Results}\label{results}

\subsection{Accuracy of suggesting road categories}

Table \ref{tab-result} lists the correctness of ChatGPT suggested road categories based on two different methods. First, we consider historical \texttt{"highway"} values of an OSM road. A suggestion was considered correct if the current or any previous versions of the corresponding OSM highway value matched the suggested tag of ChatGPT. This step takes into account differences in how individual mappers may perceive road features (e.g. primary vs. secondary). The second method is based on semantic road categories listed in Table \ref{tab1}. Considering groups of roads as opposed to individual \texttt{"highway"} values mitigates the fact that OSM tagging often follows administrative roles that are difficult to infer from photographs. Table \ref{tab-result} reports the accuracy of individual analysts across the four scenarios as well as the average correctness for analysts (values on bottom) and scenarios (values in different rows).

%\begin{adjustwidth}{-2.5 cm}{-2.5 cm}\centering\begin{threeparttable}[!htb]...\end{threeparttable}\end{adjustwidth}
\begin{table}[!htp]\centering
\caption{Accuracy score of OSM tags suggested by ChatGPT. (LC = Locational context, OD = Object detection, OD + LC = Object Detection + Locational context)}\label{tab-result}
\scriptsize
\begin{tabular}{l|r|r|r|r|rr}\toprule
\multicolumn{6}{c}{\textbf{Based on historical \texttt{"highway"} values}} \\\cmidrule{1-6}
\textbf{Scenario} & \textbf{BLIP-2} & \textbf{Analyst \#1} & \textbf{Analyst \#2} & \textbf{Avg. correct [\%]} &\textbf{\% change } \\\midrule
\hline
\textbf{Baseline} &23.4 &37.2 &31.9 &30.8 &- \\
\textbf{LC} &24.5 &46.8 &34.0 &35.1 &+4.3 \\
\textbf{OD} &27.7 &47.9 &39.4 &38.3 &+7.5 \\
\textbf{OD + LC} &30.9 &47.9 &50.0 &42.9 &+12.1 \\
\hline
\textbf{Avg. correct [\%]} &26.6 &45.0 &38.8 &\multicolumn{2}{c}{} \\
\hline
 \\
\multicolumn{6}{c}{\textbf{Based on semantic road categories}} \\
\textbf{Scenario} & \textbf{BLIP-2} & \textbf{Analyst \#1} & \textbf{Analyst \#2} &\textbf{Avg. correct [\%]} &\textbf{\% change } \\\midrule
\hline
\textbf{Baseline} &25.5 &54.3 &41.5 &40.4 &- \\
\textbf{LC} &35.1 &64.9 &45.7 &48.6 &+8.2 \\
\textbf{OD} &29.8 &63.8 &60.6 &51.4 &+11.0 \\
\textbf{OD + LC} &43.6 &66.0 &70.2 &59.9 &+19.5 \\
\hline
\textbf{Avg. correct [\%]} &33.5 &62.3 &54.5 &\multicolumn{2}{c}{} \\
\bottomrule
\end{tabular}
\end{table}

BLIP-2 achieved the lowest accuracy among the three analysts, followed by Analysts \#2 and \#1 respectively. This resembles the level of detail analysts described photographs with, which suggests that in general, providing more detailed image captions may lead to more accurate tag suggestions by ChatGPT. On average, this method increased the accuracy by up to 28.8\% between BLIP-2 (lowest detail) and Analyst \#2 (highest detail).

This is further supported by the average accuracy achieved in different scenarios. The baseline scenario, which used prompts purely based on the visual description of street-level photographs, achieved a suggestion accuracy of 30-40\% on average from the three analysts. Providing additional context in different scenarios increased this accuracy. Additional location context, i.e. specifying that the roads are located near Downtown Miami (\texttt{LC scenario}) increased suggestion accuracy by 4.3-8.2\% on average, depending on the evaluation method. This can potentially be explained by regional differences in OSM tagging practices which are usually determined by local communities. In this scenario, it is possible that the AI model considered these regional differences when suggesting \texttt{"highway"} tags. Providing a list of objects detected in the source photographs (\texttt{OD scenario}) increased the average suggestion accuracy by 7.5 - 11.0\% compared to the baseline scenario. A potential explanation for this is that objects found on and near roads provide important details that help refine the category of roads. Finally, combining additional locational and object detection contexts (\texttt{OD + LC scenario}) with the description of photographs by analysts increases suggestion accuracy by 12.1 - 19.5\% on average. It is important to mention that these improvements are observed across all analysts.

\subsection{Additional tag suggestions}

In addition to the main \texttt{"highway"} category, information about additional characteristics of roads can also be recorded in OSM. To assess such a scenario, we analyze the \texttt{"lit"} tag, which indicates the presence of lighting on a particular road segment. The \texttt{"lit"} tag is set to \texttt{"yes"} if there are lights installed along the roadway. One question explicitly asked analysts whether street lights are visible on street-level photographs. In addition, street lights are a potential object category in Mapillary detections. For the following analysis, we consider the \texttt{Object Detection enhanced scenario}. The original dataset contains 24 roadways with \texttt{"lit"="yes"}.

Table \ref{tab-lit} shows that ChatGPT correctly suggested the presence of \texttt{"lit"} tag  between 63\% (BLIP-2) and 92\% (Analyst \#2) of the existing cases. ChatGPT suggested the use of the \texttt{"lit"} tag for an additional 44 - 61 features that are potentially missing from OSM. Among these features, 59 have been suggested based on prompts from at least two analysts, and 36 have been suggested based on all three analysts.

\begin{table}[!ht]
    \caption{ChatGPT suggestions of the \texttt{"lit"} tag.}\label{tab-lit}
    \centering
    \begin{tabular}{l|l|l|l}
    \hline
        ~ & \textbf{BLIP-2} & \textbf{Analyst \#1} & \textbf{Analyst \#2} \\ \hline
        \textbf{Correctly tagged}  & 15 (63\%) & 20 (83\%) & 22 (92\%) \\ \hline
        \textbf{Additional} & 58 & 61 & 44 \\ \hline
    \end{tabular}
\end{table}

\subsection{Limitations of the results}
One limitation of this study is that we conducted the experiment on a small, geographically limited sample size. The implications of these are well studied in the GIScience literature, such as the uneven coverage of street-level photographs \cite{juhasz_user_2016} and the heterogeneity of OSM tagging \cite{girres_quality_2010}, which could limit the adaptability of our method to different areas. The other group of limitations is largely related to open problems in Computer Science, such as the so-called ``hallucinations'' of generative AI, which results in content that is false or factually incorrect \cite{bang_multitask_2023}, as well as the non-deterministic nature of ChatGPT's answers \cite{wang_can_2023}. 

\section{Summary and discussion}

The study employs ChatGPT as a mapping assistant to propose OSM road tags based on textual descriptions of street-level photos. It leverages freely available geospatial data (OSM, Mapillary) and off-the-shelf models (\texttt{GPT-3.5-turbo}, BLIP-2) to enhance collaborative mapping. ChatGPT accurately suggests OSM \texttt{"highway"} values from text in 39-45\% cases, rising to 55-62\% for semantic road categories. Substituting human analysts with BLIP-2 yielded less accurate results (27-34\%), potentially due to short captions. The study explores prompt engineering and context addition for improved accuracy. Providing road location raised accuracy by 4-8\%, object lists boosted it by 8-11\%, and combining both enhanced it by 12-20\%. The study also found that increasing the level of detail with which a road scene is described in a prompt increases the accuracy of the suggested highway tags. However, it is expected that there are limitations of this method in terms of the accuracy that can be theoretically achieved. Future research will expand OSM sample size and incorporate refined traffic-related data (e.g. speed limits, turn restrictions).

Although experiments like this are useful initial steps, we urge the GIScience community to go beyond simply applying AI in geographic contexts and focus on synergistic research that advances both the spatial sciences and AI research (see, e.g. \cite{li_blip_2022,janowicz2023philosophical}). There are multiple potential extension of this work along this idea that go beyond the case study presented in this paper. For example, the exploration of a multimodal conversational AI agent for spatial data science is a promising research direction. In theory, incorporating a spatial understanding component in a multimodal AI system allows it to comprehend and analyze geospatial data. This could result in a method for the AI to interpret and interact with geospatial data, similar to how BLIP-2 enables language models to understand images. Future research should focus on exploring the potential of this integration and on deepening our understanding of the theoretical and practical aspects of this fusion. This, in turn will advance the field of research and lay the groundwork for future innovations in comprehensive multimodal AI systems for geospatial science. 
 
\subsubsection{Data availability} The dataset supporting findings presented in this paper is freely available at: \url{https://doi.org/10.17605/OSF.IO/M9RSG}.

\subsubsection{Acknowledgements} Authors would like to thank Mei Hamaguchi and Flora Beleznay for providing image captions and visual Q\&A.

\bibliographystyle{splncs04.bst}
\bibliography{references}

\end{document}